# A newly developed 10kA-level HTS conductor: innovative tenon-mortise-based modularized conductor (TMMC) based on China ancient architecture

Jinxing Zheng[1,*], Yuan Cheng[1,2], Lei Wang[1], Fei Liu[1], Haiyang liu[1], Ming Li[1], and Lei Zhu[1]

[1] Hefei Institutes of Physical Science. Chinese Academy of Sciences, Hefei230031.China
[2] University of Science and Technology of China, Hefei 230026, China

E-mail: jxzheng@ipp.ac.cn

**Abstract**

We propose a new type of high-temperature superconducting (HTS) conductor concept: modularized conductors (MC) connected by Chinese traditional tenon-mortise (TM) connection structure, reffered as TMMC. The conductor consists of multiple concentric round sub-conductors with slots for stacking REBCO tapes. Innovatively, the REBCO stacks in the adjacent sub-conductors are arranged with the fully-misaligned configuration to enhance the critical current's isotropy with respect to magnetic field and reduce ac loss. For example, the angle between the adjacent stacks in the two adjacent sub-conductors is 45º if each sub-conductor contains 4 REBCO stacks. In order to construct the fully-misaligned configuration, the sub-conductors are designed with two open half-circular formers and connected by tenon-mortise structure which makes the conductor modulrized and simply to assembly and disassembly. Based on the design concept, a prototype conductor containing 160 REBCO tapes distributed in the four concentric sub-conductors is fabricated. The conductor's measured critical current is 13.69 kA at 77 K and sefl-field, which is consistent to the simulaiton result. In order to further improve the TMMC's engineering critical current density ($J_{ce}$) and bending performance, we propose two enhancement approaches: reducing the former's thickness and re-arrange stacks in the outer sub-conductors. With the enhancements, both TMMC's radius and $J_{ce}$ are comparable to the existing slotted-core conductor. The study shows the TMMC's advantages of non-twisted structures, easy assembly, high-current carrying and low ac losses, which makes it promising for constructing large-scale scientific devices.

## 1. Introduction

The second-generation (2G) high-temperature-superconducting (HTS) materials, namely the rare-earth-barium-copper-oxide (REBCO) tapes, are promising in manufacturing both small-aperture high-static-field (>30 T) solenoid coils [1,2] and giant magnets under transient electromagnetic environment for large-scale scientific devices, such as fusion reactors [3] due to the outstanding current-carrying capacity and good mechanical stability [4]. The small-aperture high-static-field coils possess relatively simple structures consisting of pancakes which are wound with one single tape. However, the single-tape conductor results in high inductance, which makes the coil carry currents of no more than a few hundred amperes under very slow current-changing rate [5]. To achieve transient high field in a meter-class large-scale device, the conductors require low inductance and kA-class current. Therefore, the first task and challenge of employing REBCO tapes in large-scale scientific devices is to manufacture a high-current cable/conductor.

Several conductor/cable designs have been proposed using REBCO tapes, and most of them present the twisted or transposed concepts, constituting cable-in-conduit conductors (CICC) inspired by the low-temperature-superconducting (LTS) cables and conductors [6], such as Roebel assembled coated conductor cable (RACC) [7], a Roebel-structure transposed cable (X cable) [8], twisted-stacked-tape cable, TSTC [9], Quasi-round TSTC (QisTAc) [10], slotted-core cable-in-conduit conductors [11], conductor on round core [12], conductor-in-tube cables (CIT) [13], round soldered-twisted stacked [14], CrossConductor [15], and twisted Quasi-Isotropic strand [16]. The main reason for manufacturing LTS conductors with twisting is to avoid instabilities of LTS materials. Instead, the stability of HTS is much higher than LTS, HTS conductors can be made of non-twisted tapes, such as non-twisted stacked tapes assembled in rigid structure [17] and Quasi-Isotropic strand [18]. Nevertheless, twisting, as an approach to reducing inductance mismatch, improving even current distribution in tapes and reducing AC loss, is employed to enhance the performance of HTS conductors [19]. However, recent studies showed that non-twisted HTS cables' inductance variation is typically less than 2%, which results in slight effect of uneven current distribution, and the reduction by twisting is less than 36% in terms of ac loss for HTS cables [6]. Therefore, the benefits of performance enhancement brought by the twisting is limited with comparison to the much easier fabrication process of non-twisted conductors.

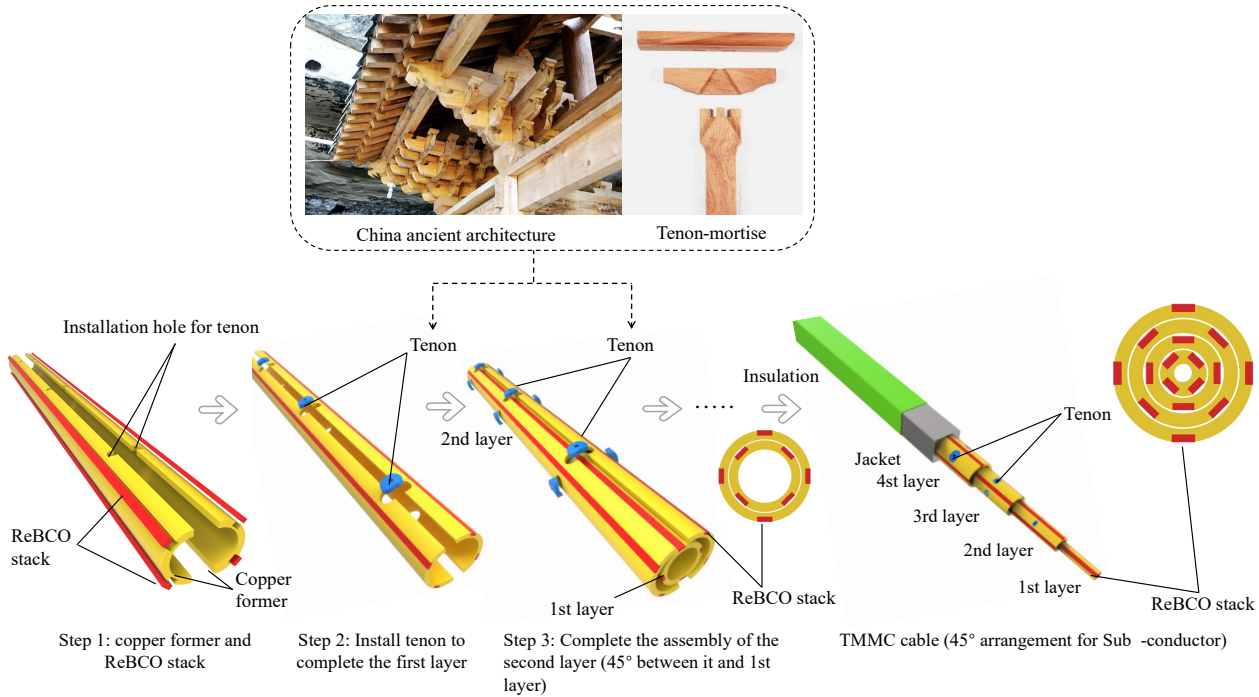

**Figure 1.** the structure of the modularized conductor connected by tenon-mortise structure

Meanwhile, considering of the highly anisotropic properties and remarkable ac loss effects of REBCO tapes, optimized structures are also employed in the non-twisted HTS conductors. For example, tapes are arranged vertically and horizontally in the Quasi-Isotropic strand to reduce the anisotropic values of critical current with repect to magnetic field [18]. It is meaningful to explore more arrangement configurtions of REBCO stacks to enhance the performance of non-twisted conductors. But this would be restricted by soldering structure, in which slots for stacking REBCO tapes are limited to unidirectional directions. Besides, the conductors are difficult to disassembly and repair when the REBCO stacks are soldered with the conductor's former as a while one. Modularized conductors without soldering will be a promising concept to solve the issues, and this type of conductor has not been developed before.

In this paper, we propose a new type of HTS conductor concept: modularized conductors (MC) which are assembled without soldering but Chinese traditional tenon-mortise (TM) connection structure, therefore the new conductor is referred as TMMC. The article is structured as follows. Section 2 introduces the idea and concept design of the modularized conductor baded on tenon-mortise connection. The detailed optimization process of designing the stacks' configuration is presented in Section 3. A prototype conductor with 4 concentric sub-conductors is fabricated and its critical current at 77 K and self-field is measured and simulated in Section 4. Section 5 gives the discussions about enhancing the TMMC's engineering critical current density and reducing ac losses further.

## 2. Modularized conductor based on tenon-mortise structure

Figure 1 illustrates the structure and assembly process of the proposed Tenon-Mortise Modularized Conductor (TMMC) composing of multiple concentric round sub-conductors with varying diameters. The innermost central channel (diameter 6 mm) is used for coolant flow. Each sub-conductor consists of slotted metal formers with REBCO stacks in the slots. The surfaces of the REBCO tapes are distributed along the circumferences of the formers to reduce the influence of perpendicular magnetic fields on critical current and ac loss. Innovatively, the two adjacent sub-conductors are assembled with the fully-disaligned configuration, which is determined by the angle between the centers of two REBCO stakcs in the adjacent sub-conductors. For instance, if each sub-conductor holds 4 REBCO stacks, the max misalignment angle is 45 degrees, as demonstrated in figure 2. This fully-misaligned configuration is quite different from the previous conductor layout with slotted former/core, in which the slot number is fewer and tapes in each slot are stacked in unidirection direction [11]. The new configuration significantly improves critical current isotropy with respect to magnetic field and reduces ac losses, which has been verified based on simulations in Section 3.

A challenge is to maintain the misaligned configurations of REBCO stacks in the different sub-conductors. In order to facilitate manufacturing long conductors, we design the sub-conductors consisting two open half-cirlces, which are connected as a complete round sub-conductor by the tenon-mortise structure inspired by the traditional Chinese connection method used in architecture, furniture, and instruments. The connection structure bascially includes two components: convex (tenon) and concave (mortise) parts, which have many variants in practical. In our design, the swallow-tailed tenons and mortises are used to joint the two open half-circles, as shown in figure 1. The tenon-mortise joints in the outer sub-conductor align with the slots of REBCO stacks in the adjacent inner sub-conductor and the misaglignment configuration generates. Additional shallow grooves perpendicular to the tape stacking slots are cutted, therefore, the swallow-tailed tenons can stabilize the REBCO stacks and prevent sub-conductor's movement in axial and circumferential directions when they are inserted into the mortises and grooves.

Since the assembly of sub-conductors is independent from each other, the conductor is easily fabricated to multiple layers to carry high currents, which makes the first successful development of a modularized REBCO conductor. However, the bending performance and engineering current density of the conductor will be decreased because of the multiple-layer structures which results in larger diameters. To solve the issues, one can minimise the thicknesses of the formers, and besides re-arrange the REBCO stacks in the outer sub-conductors, which are discussed in Section 5.

## 3. The optimization process for TMMC

### 3.1 Effects of stack's arrangement on the conductor's ac loss

Firstly, we design a simplified conductor with 4 stacks and 2-layer tapes per stack to investigate the effects of tape's arrangement in different layers on the conductor's ac losses. Different tape-layout structures are characterized by the misalignment angle ($\alpha$) between the centre of the two layers of tapes, whose maximum angle is 45 degrees in this case from figure 2.

Different configurations with various angles are simulated by $H$ formulation and finite-element method [20]. Considering magnetic field ($H$) as the dependent variable, the Maxwell's equations can be solved as:

$$J = \nabla \times H \tag{1}$$
$$E = \rho \cdot J \tag{2}$$
$$\rho = \frac{E_c}{J_c(B)} \left| \frac{J}{J_c(B)} \right|^{n-1} \tag{3}$$
$$\nabla \times E = -\mu_0 \frac{\partial H}{\partial t} \tag{4}$$

where $E$ is the induced electric field and $J$ is the transporting current density. $J_c$ is the tape's critical current density. $E_c$=1E-4 V/m is the value of voltage criterion for critical current measurement, and $n$=25 is defined from DC measurement of the superconductor's highly non-linear $I$-$V$ characteristic. The Kim-like model [21] is used to describe the anisotropic dependence of the tape's critical current density on magnetic field, and the fitting equation of $J_c$ at 4.2 K is defined as:

$$J_c(B) = \frac{J_{c0}}{\left(1 + \frac{\sqrt{k^2 B_\parallel^2 + B_\perp^2}}{B_0}\right)^\alpha} \tag{5}$$

where $J_{c0}$=3.74e11 A/m$^2$, $B_0$=1.67279 T, $k$=9.13e-3, and $\alpha$=0.76939 can be found in [22]. Integral method is employed to impose current in the conductors and implemented by global constraint in the finite-element software.

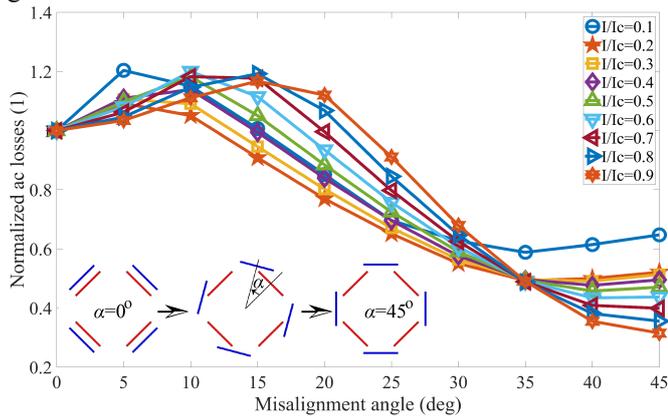

**Figure 2.** Change of two-layer conductor's ac loss with the misalignment angle between the adjacent tapes.

Normalized AC losses are calculated by dividing values at different angles by those at α=0 degree, representing the traditional unidirectional tape-stacking configuration. In the following, we refer the slotted-core conductor with groves in a single-layer former as the unidirectional-stack conductor. Figure 2 demonstrates that as the angle increases, AC losses first rise and then decline. Comparing losses at α=0 and α=45 degrees, the latter is about 30%-50% of the former, indicating significant AC loss reduction with fully-misaligned configurations. It is important to note that AC losses show minor increases at low currents beyond a 35-degree misalignment angle but diminish at higher currents. To facilitate manufacturing, the fully-misaligned configuration with a fixed angle is chosen.

A question arises when stacking more-layer REBCO tapes. Take the example of a 3-layer conductor: should the third layer be stacked with the maximum misalignment angle calculated from the adjacent layers (e.g. α=45 degrees with 4 tapes in each layer), or should all three layers be considered as a whole to reconstruct a fully-misaligned configuration (e.g. α=30 degrees), as shown in figure 3. The comparison indicates that the configuration with the maximum misalignment angle derived from the two adjacent layers results in lower AC losses.

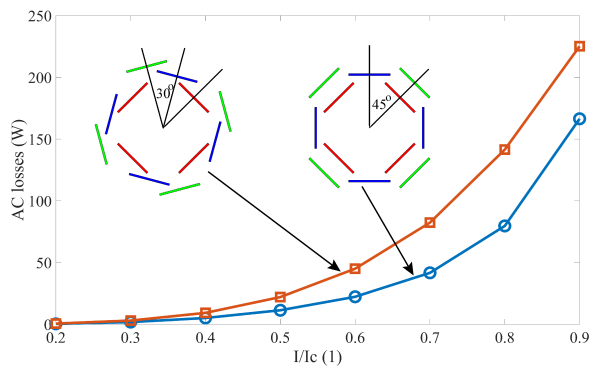

**Figure 3.** Comparisons of ac loss in 3-layer conductors with different stacking configurations.

These studies indicate that REBCO-based conductors' ac losses are significantly affected by the stacking configurations. We can reduce losses by re-arranging tapes with the maximum misalignment angle between the tapes in the adjacent layers.

### 3.2 Layer number affecting the conductor's critical current and ac loss

Above studies demonstrate that minimal AC losses occur when adjacent layers of tapes are distributed at the maximum misalignment angle. In practical high-current conductor manufacturing with many tape in one stack, it is not possible to misalign every single tape as one layer. Consequently, only a limited number of layers can be misaligned, with tapes in each layer still stacked unidirectionally.

In this part, we investigate the impact of the layer numbers on the conductor's electromagnetic properties. The conductor contains 160 REBCO tapes distributed in the former with 4 slots around a circular cooling channel. In the conventional stacked conductor, each stack comprises 40 REBCO tapes, marked by 1 layer in figure 4.

Following the misaligned configuration, we construct conductors with 2, 3, and 4 layers, with adjacent layers aligned at 45 degrees. Initially, we examine the critical currents of these four conductors under an external magnetic field of 20 T at 4.2 K using the load-line method. As the critical current of REBCO tapes strongly depends on the angle between the tapes and the magnetic field, we vary this angle from 0 to 45 degrees to assess the effect of layer numbers on the anisotropy of critical currents with respect to the magnetic field. As presented in figure 4, the critical current isotropy in even layers is superior to that in odd layers. In odd layers, isotropy improves with an increasing number of layers, while isotropy in 2 and 4 layers remains similar.

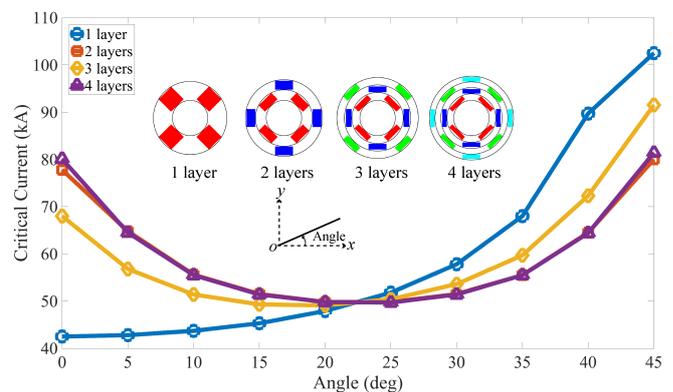

**Figure 4.** Variations of conductors' critical currents with respect to external magnetic field at 20 T and 4.2 K.

Meanwhile, we calculate the conductors' ac losses with peak currents of 8 kA at self-field. The accumulated ac losses in one cycle are presented in figure 5, which shows that ac losses get smaller with more layers, and the reduction is limited as layers are built more. Thus, we design a 4-layer conductor in the study to balance the ac loss reduction and fabrication process.

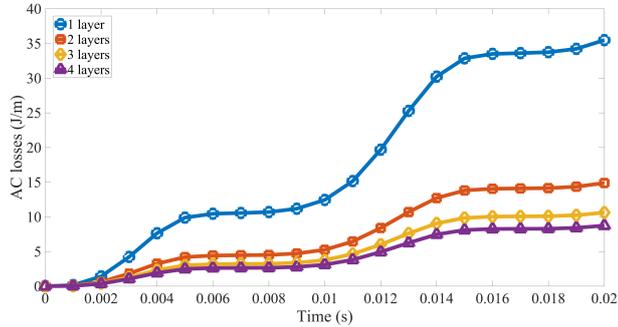

**Figure 5.** Variations of conductors' ac losses at self-field.

## 4. Fabrication and performance test of the prototype TMMC

### 4.1 Parameters of the prototype TMMC

We have manufactured a prototype TMMC according to the above design concept, which consists of 4 concentric sub-conductors (referred as layer 1 to 4 from inside out) accommodating a total of 160 REBCO tapes. Each sub-conductor has 4 slots with 10 tapes in each, as shown in figure 6. Geometric parameters of the prototpye TMMC are give in table 1.

The REBCO tapes are produced by Shanghai Superconductor with 4 mm width and 144 A self-field critical current at 77 K. The tape's anisotropic dependence of critical current on magnetic field is measured and calculated in simulations by interpolation function of the data. The numerical models for calculating conductor's critical current are implemented based on $H$ formulation, in which the superconducting region is considered as a whole to carry current so that the non-uniform current repartition among tapes can be presented. Besides, the MAX criterion is selected as the criteria determining conductor's critical current, which means that when the voltage drop per unit length reaches the critical value $E_c$ (1μV/cm) in at least one tape, the imposed current is considered as the conductor's critical current [23]. According to the simulations, the estimated critical currents of the prototype 1-layer and 4-layer TMMCs are 4.072 kA and 15.328 kA, respectively, at 77 K and self-field.

### 4.2 Ic measurement for the prototype TMMC

REBCO tapes are longer than the formers of the sub-conductors in the cable fabrication [24][25], so that the tapes and the ends of the prototype TMMC can be soldered with two copper joints, shown in figure 6. It is imperative to observe that the temperature during joint welding must not surpass 200°C [26]. Currents are injected into REBCO tapes through the joints. In the experiment, the prototype TMMC is immersed in a liquid nitrogen (LN$_2$) tank, and the copper joints connect with the direct current (DC) source. The temperature of the conductor is considered to be 77 K when liquid nitrogen ceases to boil [27][28].

Table 1. Geometric parameters of the prototype TMMC.

| Parameters | Value |
|---|---|
| Sub-conductor number | 4 |
| Slot number in each sub-conductor | 4 |
| Tape number in each slot | 10 |
| The depth of each slot | 2 mm |
| Cable length | 800 mm |
| Outer and inner diameters of layer 1 | 10 mm/ 4 mm |
| Outer and inner diameters of layer 2 | 17 mm/ 11 mm |
| Outer and inner diameters of layer 3 | 24 mm/ 18mm |
| Outer and inner diameters of layer 4 | 31 mm/ 25 mm |

Firstly, we measure the critical current of the prototype TMMC with only 1-layer sub-conductor contining 40 tapes, whose data is depicted in the figure 7. The measured value is 4012 A at 77 K and self-field, which does not differ significantly from the predicted one.

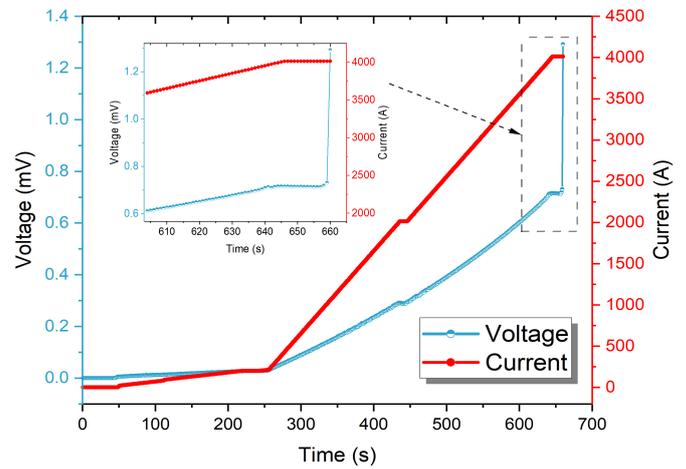

**Figure 7.** $I_c$ measurement for one-layer TMMC.

Then, we implement the $I_c$ measurement of the complete prototype TMMC with 4 concentric sub-conductors. The experimental equipments and parameters are given in figure 8a-b and table 2.

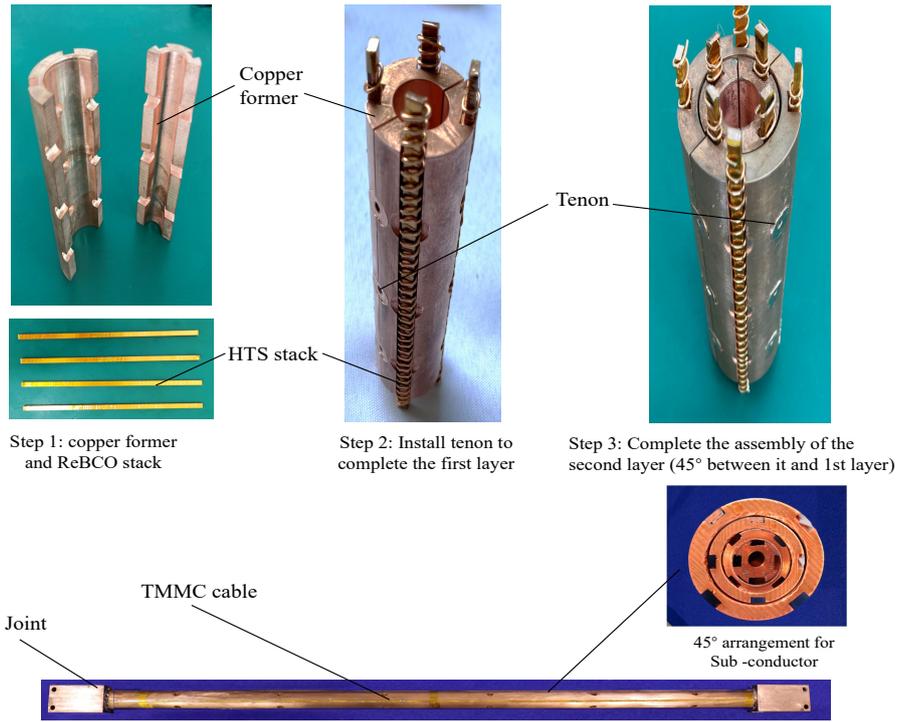

**Figure 6.** The prototype TMMC: the Tendon-tenon connection and the assemby of TMMC conductor

Table 2. Parameters of DC experiment for $I_c$ measurement.

| Parameters | Value |
| --- | --- |
| DC source | 100 kA |
| Voltage acquisition equipment | EF721 |
| Signal acquisition frequency | 10 Hz |
| d$I$/dt | 50 A/s |
| $V_{th}$ | 70 uV |
| Estimated $I_c$ | 15 kA |

In the experiment of ReBCO cable, it is customary to employ multiple voltage probes for the purpose of discerning the state of the cable [29][30]. Figure 8(c) illustrates voltage probes soldered on the TMMC: V1 and V3 for measuring conductor joint voltage, and V2, V4, and V5 for recording voltage within the superconducting segments. Since the length between V2 probes is 700 mm, the threshold voltage ($V_{th}$) for determining critical current is set at 70 μV, taking into account base voltage and noise removal.

Figures 9(a) and 9(b) present data of currents and voltages during the measurement processes. The resistance of the joints on both sides is 110 nΩ and 140 nΩ, respectively. The voltages are significantly increased when the TMMC transits from superconducting to non-superconducting states. The measured critical currents ($I_c$) of the 4-layer prototype TMMC at 77 K under self-field and additional external 0.7 T are 13.69 kA and 10.58kA, respectively.

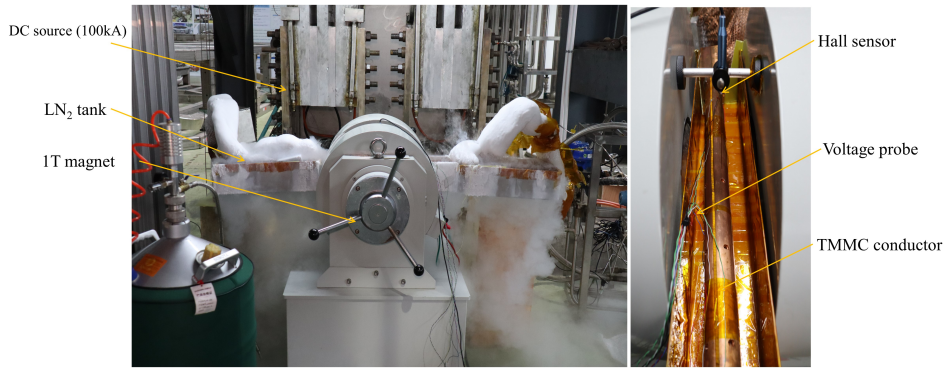

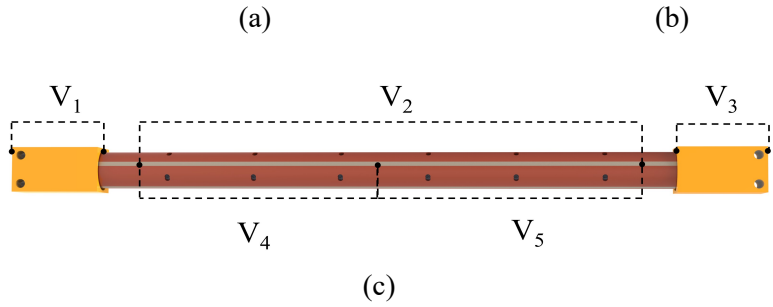

**Figure 8.** $I_c$ measurement for TMMC: (a) DC source and the prototype TMMC; (b) Hall sensor and TMMC; (c) the voltage probes for conductor.

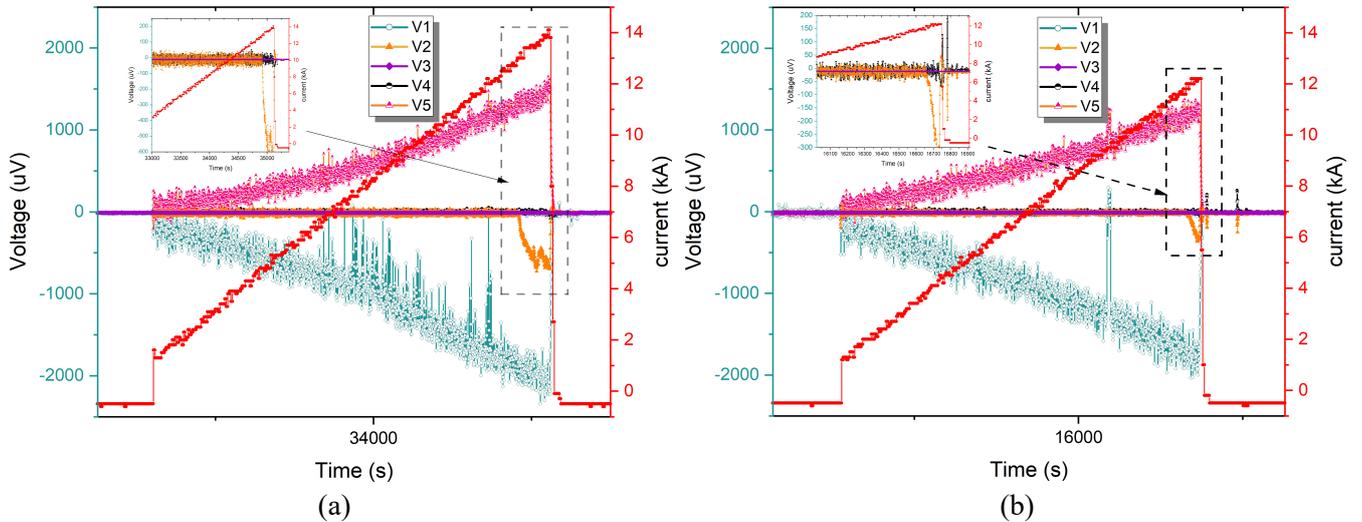

**Figure 9.** $I_c$ of the prototype TMMC at 77 K: (a) $I_c$ under self-field, (b) $I_c$ under self-field and external 0.7 T.

## 5. Discussions

### 5.1 Enhancing conductor's critical current density

Although the structure of concentric sub-conductors connected by tenon-mortise joints makes the conductor modularized and easy to be assembled, the multiple formers influence the conductor's engineering critical current density compared with the traditional stacked-tape conductors. And the engineering critical current density is 55 A/mm² in the slotted-core conductor [11]. One can increase the engineering critical current density by reducing the former's thickness, while this is limited considering of the former's mechanical strength. For example, the critical current and its density of the originally manufactured TMMC with each-layer former's thickness of 4.5 mm are predicted to be 15.33 kA and 11.06 A/mm², respectively. And if we reduce the former's thickness to 2.0 mm, the two parameters are 13.93 kA and 36.64 A/mm², respectively.

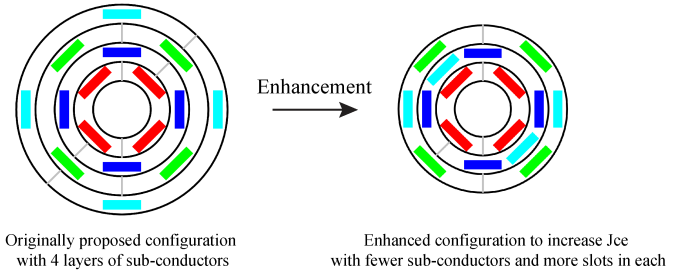

**Figure 10.** Enhancing the conductor's critical current density with less sub-conductors. The gray lines in every sub-conductor represent the joints of the two half-circles.

Here, we discuss another approach that is cutting more slots in the outer concentric formers based on the above proposed structures so that fewer sub-conductors are needed, as shown in figure 10. Taking the 2mm-thickness former as an example, we re-arrange the 4 stacks in the layer 4 into the layer 2 and 3. After the enhancement, the conductor's critical current at 77 K, self-field and its engineering critical current density are 13.51 kA and 53.09 A/mm², which is comparable to the that of the unidirectional-stack conductor. Besides, the diameter of the all formers in the enhanced configuration is 18 mm, close to that of the slotted-core conductor (diameter 19 mm) [11], therefore, bending the TMMC to manufacture magnets is feasible. In the current design, the enhanced TMMC's engineering critical current density is similar to that of the slotted-core conductor, but it can be significantly improved with the approach of re-arranging the stacks in the outer sub-conductors compared to the latter when the conductor contains more tapes. Because more slots can be cut in the larger-radius formers of TMMC to contain more tapes, while the tapes can be only stacked in unidirectional direction in the traditional slotted-core conductor.

*5.2 AC loss comparison study*

AC losses of the prototype TMMC with different radii are calculated under the same engineering current density, and compared with the traditional unidirectional stacks. The normalized current density distributions and ac losses of different conductors are shown in figure 11 and 12. The conductors with varying thicknesses of formers carry different transporting currents, while keep same engineering current densities and current ramping rate (500 A/s). Triangle-wave currents are applied in the whole conductor therefore, non-even current distributions are considered. From figure 11, it can be seen that the imposed currents are mostly distributed in the outmost tapes, and the current penetration becomes shallower with the TMMC's former reducing because of the increased engineering critical current density. Correspondingly, the TMMC's ac loss is significantly influenced by the conductor's radius too, shown in figure 12. With the former of sub-conductor decreasing by 1 mm, the peak loss of TMMC during the initial current-imposing period reduces by more than 75%. When the former's thickness reduces to 2 mm, the ac loss of TMMC is close to that of the traditionally unidirectional-stack conductor. And among all the conductors, the enhanced configuration based on the 2-mm former TMMC achieves the minimum ac losses. Due to the increased engineering critical current density, the peak loss reduces by 57% and 35% compared with the non-enhanced 2mm-former TMMC and unidirectional-stack conductor, respectively.

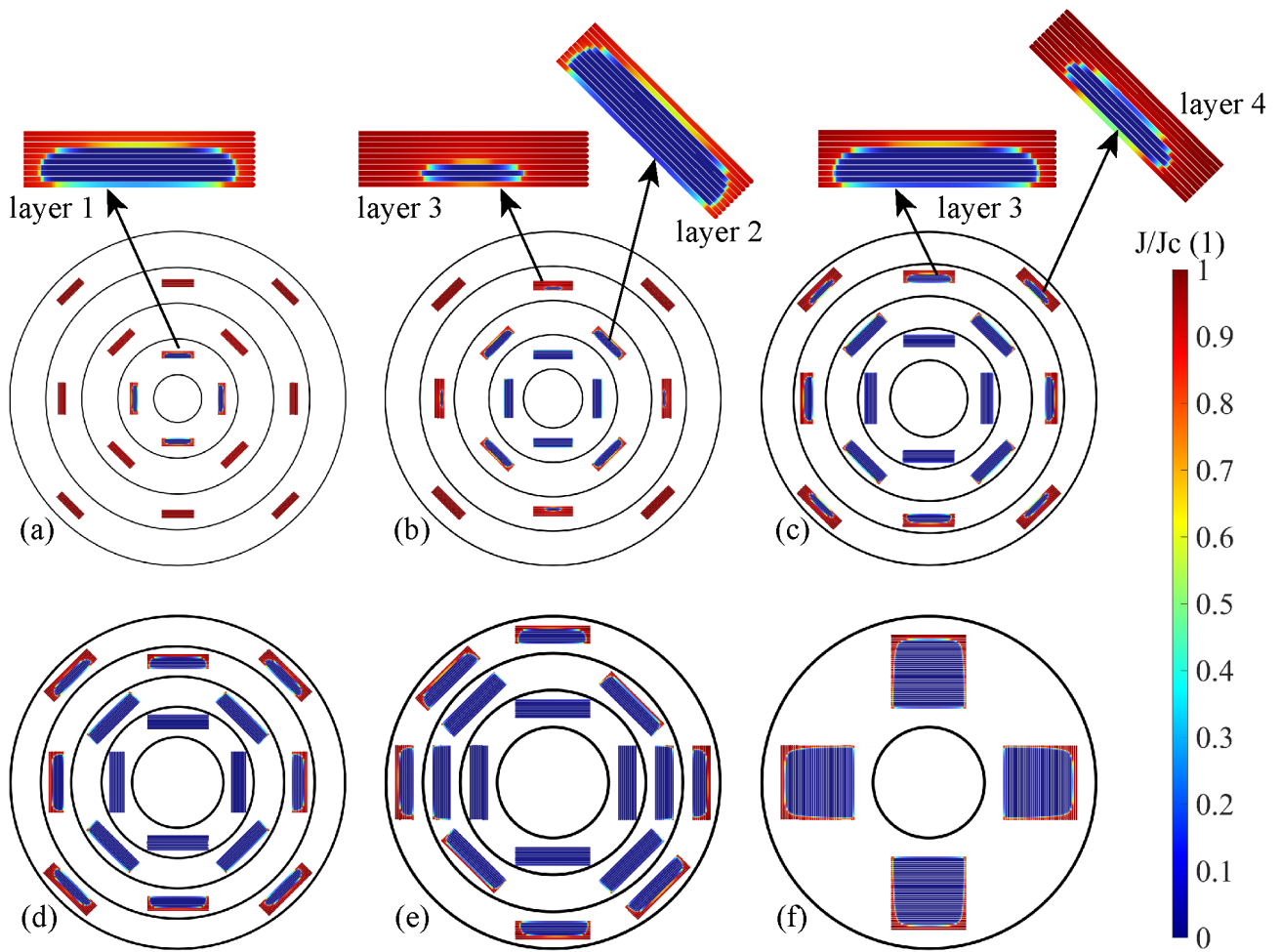

**Figure 11.** Distributions of normalized current density in conductors: (a)-(d) represent TMMCs (160 tapes) with 4.5-mm, 3.5-mm, 2.5-mm and 2-mm formers, respectively. (e) is the enhanced configuration based on the 2-mm former. (f) represents the traditionally unidirectional-stack conductor.

## 6. Conclusions

In this paper, a new kind of HTS modularized conductor (MC) concept is proposed, in which the conductor consists of multiple concentric half-circle sub-conductors connected by Chinese traditional tenon-mortise (TM) structure. Therefore, the proposed HTS conductor is referred as TMMC. Innovatively, the REBCO stacks are arranged by the fully-disalligned configurations of the adjacent sub-conductors, which benefits for improving the isotropy of conductor's critical current on magnetic field and reducing ac loss.

According to the design concept, prototype TMMCs are manufactured. The measured critical currents of 1-layer and 4-layer TMMCs are 4.012 kA and 13.69 kA, respectively, at 77 K and self-field, which are close to the predicted values of 4.072 kA and 15.328 kA.

In order to further improve the TMMC's engineering critical current density ($J_{ce}$) and bending performance, we propose two enhancment approaches to reduce TMMC's radius: reducing the former's thickness and re-arrange stacks in the outer sub-conductors. With the enhancements, the TMMC's $J_{ce}$ is increased by 4 times. The TMMC's radius, $J_{ce}$ and ac losses are comparable to the unidirectional-stack conductor, which makes it feasible to construct magnets.

With the advantages of non-twisted structures, easy assembly, high-current carrying capacity and low ac losses, the proposed tenon-mortise modularized conductor is promising for constructing large-scale scientific devices. And we are planning to implement more tests to verify the performances of TMMC and promote it to be employed more widely.

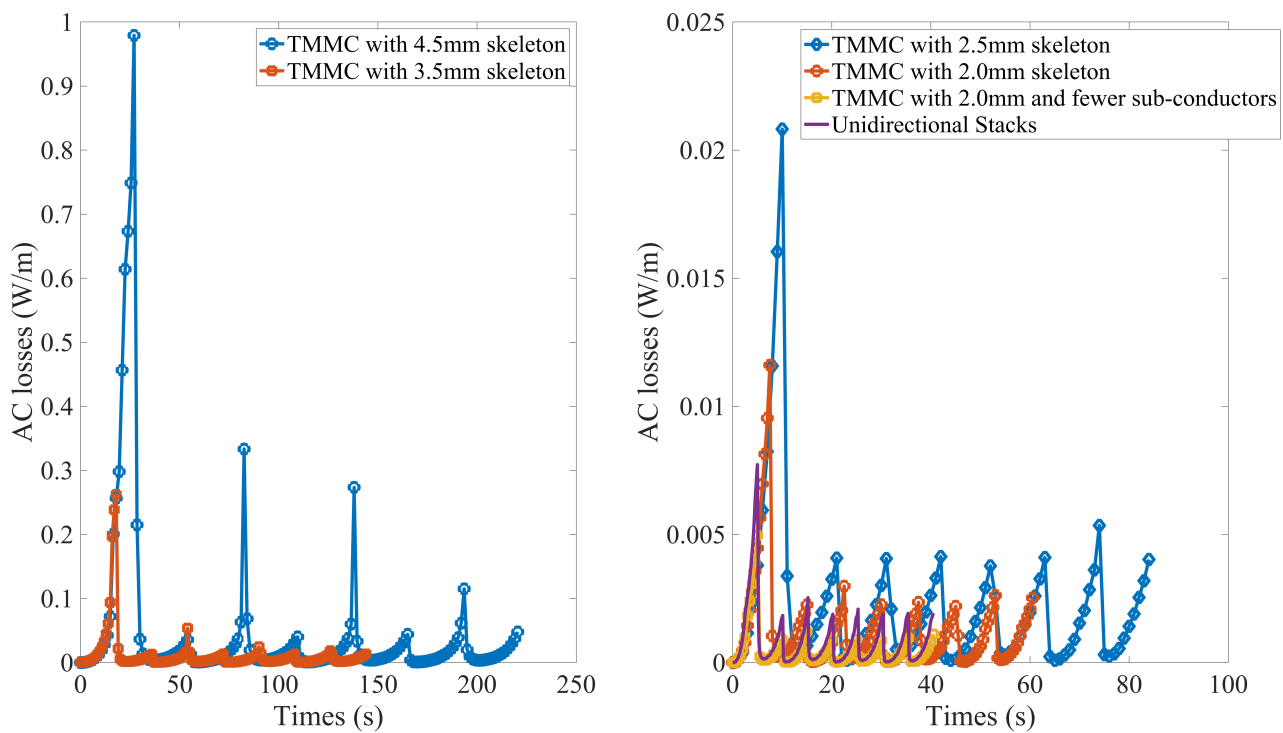

**Figure 12**. AC losses comparisons of TMMCs with different radii and traditionally unidirectional stack's conductor under the same carrying current density.

## Acknowledgements

This work was supported by the National Natural Science Foundation of China General Program 52077211, by the National Science Found for Excellent Young Scholars of China 52222701, and by the Anhui Province Key Research and Development Plan 2022i01020019.